\documentstyle{aipproc}
\begin{document}
\title{Consistent Perturbative Light Front\\ Formulation of Yang-Mills Theories}

\author{M. Morara$^*$, R. Soldati$^*$ and G. McCartor$^\dagger$}
\address{$^*$Dipartimento di Fisica "A. Righi", Universit\`a di Bologna\\
$^{\dagger}$ Department of Physics, SMU, Dallas TX}

\maketitle

\begin{abstract} 
It is shown how to obtain the consistent light front form
quantization of a non-Abelian pure Yang-Mills theory (gluondynamics) in the
framework of the standard perturbative approach. After a short review of the
previous attempts in the light cone gauge $A_-=0$, it is  explained how the
difficulties can be overcome after turning to the anti light cone gauge
$A_+=0$. In particular, the generating functional of the renormalized
Green's functions turns out to be the same as in the conventional instant
form approach, leading to the Mandelstam-Leibbrandt prescription for the
free gluon propagator.
\end{abstract}

\section*{Light Front Form for Free Fields}
The light front form ({\bf LFF}) formulation\footnote{$\underline{Notations}$:
$$
v^\mu\equiv (v^+,v^-,v^\perp),\
v^\pm =
{v^0\pm v^3\over \sqrt 2},\
v^\perp=(v^1,v^2)=v^\alpha.
$$}
of field theories
\cite{1.} is given in terms of
the evolution parameter $x^+$, the light front "time", and
the "volume" coordinates ${\bf x}=(x^-,x^\perp)$, which label the light
front
"space".
The LFF formulation of gauge theories in the light cone gauge
$
A_-=n^\mu A_\mu=0,
$
is known since almost thirty years \cite{2.}. This original formulation involves
only physical degrees of freedom and, at the perturbative level, it
unavoidably leads to the Cauchy's principal value (CPV) prescription
for the non-covariant spurious singularity of the free gauge boson
propagator:
namely,
$$
D^+_{\mu\nu}(x)=i\int{d^4 k\over (2\pi)^4}
{e^{ikx}\over k^2+i\epsilon}\left\{
-g_{\mu\nu}+{n_\mu k_\nu + n_\nu k_\mu\over [nk]_{\rm CPV}}\right\},
\label{1}
$$
where
$$
{1\over [nk]^2_{\rm CPV}}\equiv {\cal S}^\prime-\lim_{\varepsilon\to
0}\ {nk\over nk^2  + \varepsilon^2}\ ,
\label{2}
$$
the limit being understood in the tempered distributions' topology.

Unfortunately the CPV prescription so roughly violates
power counting and causality, that it even fails to reproduce the correct
one
loop beta function in the non-Abelian case \cite{3.}. It follows therefrom that
the original light front form perturbation theory in the light cone gauge is
inconsistent.

On the other hand, it turns out that the usual instant form ({\bf IF})
formulation, which is in terms of
the evolution parameter $x^0=ct$, the ordinary time, and
volume coordinates $\vec x$ for the ordinary space,
is fully consistent in perturbative approaches to
gauge theories \cite{4.}, both
in the light-cone gauge $A_0-A_3=0$
as well as
in the anti light cone gauge $A_0+A_3=0$.

This {\bf IF} formulation necessarily
involves extra unphysical degrees of freedom (ghosts)
and canonical quantization necessarily leads
to the Mandelstam-Leibbrandt ({\bf ML})
tempered distribution \cite{5.} to regulate the non-covariant singularity in the
free
gauge boson propagator, i.e.,
$$
{1\over [k_0\mp k_3]_{\rm ML}}\equiv {\cal S}^\prime-\lim_{\varepsilon\to
0}\ {k_0\pm k_3\over (k_0\pm k_3)(k_0\mp  k_3)+ i\varepsilon}.
\label{3}
$$

The {\bf ML} prescription guarantees
power counting and causality
and in so doing the {\bf IF} perturbative formulation does fulfill
renormalizability, unitarity and
covariance of the formal {\bf S}-matrix elements.
The natural question arises:
is it possible to find some {\bf LFF} formulation
which reproduces those remarkable results?
For Quantum Electrodynamics (the Abelian theory) the answer is  yes \cite{6.},
provided we set up canonical light front form quantization in the Weyl's
gauge
$A_+=0$.
In order to obtain some consistent {\bf LFF} perturbation theory we have to
find some canonical {\bf LFF} framework leading to the {\bf ML}
form of the free propagator.
The first attempt towards this task has been pionereed,
in the light-cone gauge, by
G. McCartor and D. G. Robertson \cite{7.}.

Their starting point is the free Lagrange density
$$
{\cal L}_{{\rm rad}}=-{1\over 4}F_{\mu\nu}F^{\mu\nu}-\Lambda n^\mu A_\mu,
\label{4}
$$
in which
$
n^\mu=(n^+,n^-,n^\perp)=(0,1,0,0).
$
After the introduction of the new field variables
$
(A_\alpha,A_+,\Lambda)\ \longmapsto\ (T_\alpha,\varphi,\lambda)
$, i.e.,
\begin{eqnarray}
& A_\alpha& = T_\alpha -\partial_\alpha \varphi,
\quad \Lambda =\partial_\perp^2\lambda,\\
& A_+& = \partial_\alpha \partial_-^{-1}* T_\alpha -
\partial_+ \varphi\ -\lambda,
\label{5}
\end{eqnarray}
the equations of motion become
$$
2\partial_+\partial_- T_\alpha =  \partial_\perp^2T_\alpha,
\quad
\partial_-\varphi=
\partial_-\lambda =0.
\label{6}
$$
We remark that $\varphi$ and $\lambda$
fulfill constraint equations and thereby
$\varphi$ and $\lambda$ can not be quantized
on the null hyperplanes at constant $x^+$.
Owing to this feature, in \cite{7.} some new {\bf LFF} quantization procedure
was suggested
involving two characteristic surfaces, i.e.,
transverse fields $T_\alpha$ are quantized
 on null hyperplanes at
equal $x^+$, whereas
longitudinal fields  $\varphi$ and $\lambda$ are quantized
on null hyperplanes
at equal $x^-$. The above quantization procedure leads to the
light front form canonical commutation relations
listed below
\begin{eqnarray}
\left[T_\alpha (x),T_\beta (y)\right]_{x^+=y^+} & =&
{\delta_{\alpha\beta}\over 2i}{\rm sgn} (x^- -y^-)
\delta^{(2)}(x^\perp-y^\perp), \label{7a}\\
\left[\varphi (x),\lambda (y)\right]_{x^-=y^-} & =&
i\delta (x^+ -y^+)\partial_\perp^{-2}*\delta^{(2)}
(x^\perp-y^\perp), \label{7b}
\end{eqnarray}
\begin{equation}
\left[T_\alpha (x),\varphi (y)\right]=
\left[T_\alpha (x),\lambda (y)\right]=
\left[\varphi(x),\varphi (y)\right]=
\left[\lambda (x),\lambda (y)\right]=0.
\label{7c}
\end{equation}
However, when we compute the {\bf LFF} ordered product
\begin{equation}
D^+_{\mu\nu}(x-y) \equiv\theta (x^+ -y^+)\left<0|A_\mu (x) A_\nu
(y)|0\right>
 +\theta (y^+ -x^+)\left<0|A_\nu (y) A_\mu (x)|0\right>,
\label{8}
\end{equation}
the {\bf LFF} quantization scheme of Eqs.\ (\ref{7a}), (\ref{7b}) and (\ref{7c}) drives to
ill-defined
convolution products
and not
to the {\bf ML} form of the
vector boson propagator. As a consequence, it turns out that
the {\bf LFF} quantization of gauge theories in the light cone gauge
$A_-=0$ is indeed troublesome, when choosing $x^+$ as the evolution
parameter.
The simplest way to overcome the above barring is the transition to the anti
light cone gauge, or light front form Weyl's gauge,
$
n^{*\mu}A_\mu = A_+=0,\
n^{*\mu}\equiv (1,0,0,0).
$
Consider therefore the new Lagrange density for the free Maxwell's radiation
field
$$
{\cal L}_{{\rm rad}}=-{1\over 4}F_{\mu\nu}F^{\mu\nu}-\Lambda n^{*\mu} A_\mu;
\label{9}
$$
The best way to set up the light front form quantization of the above
constrained system is to follow Dirac's method \cite{8.} of canonical
quantization.
The free (unconstrained) canonical momentum is
$
\pi^- = F_{+-},
$
and we have the second class primary constraints
$
\pi^\alpha - F_{-\alpha}=0,
$
as well as the first class primary constraints
$
\pi^+ = \pi^\Lambda = 0.
$
The canonical Hamilton density is
$$
{\cal H}_{{\rm rad}}={1\over 2}\left(\pi^-\right)^2+
{1\over 4}F_{\alpha\beta}F_{\alpha\beta} -
A_+\left(\partial_\alpha\pi^\alpha +
\partial_-\pi^- - \Lambda\right),
\label{10}
$$
whence we derive the secondary constraints
$
A_+=0,\
\partial_\alpha\pi^\alpha +\partial_-\pi^- = \Lambda.
$
The full set of constraints is now second class and thereby
we can compute equal $x^+$
Dirac's brackets, whose explicit form can be found in \cite{6.}.

After introduction of the new set of variables
\begin{eqnarray}
A_\alpha &=&T_\alpha -\partial_\alpha\varphi,
\quad \pi^- =\partial_\alpha T_\alpha,\\
A_- &=& 2\partial_- \partial_\alpha \partial_\perp^{-2}*T_\alpha -
\partial_-\varphi\ -\lambda,
\label{11}
\end{eqnarray}
we obtain the genuine equations of motion for all the fields, i.e.,
$$
2\partial_-\partial_+ T_\alpha  = \partial_\perp^2 T_\alpha,\quad
\partial_+\varphi =
\partial_+\lambda=0.
\label{12}
$$
The transition to the quantum theory is achieved after replacement of the
light front form Dirac's brackets with the corresponding light front form
canonical commutation relations, which now read the same as in Eq.s\ (\ref{7a}), (\ref{7b}) and (\ref{7c}), but
for the crucial Eq.\ (\ref{7b}) which is replaced by
$$
\left[\varphi (x),\lambda (y)\right]_{x^+=y^+}  =
i\delta (x^- -y^-)\partial_\perp^{-2}*\delta^{(2)}
(x^\perp-y^\perp).
\label{13}
$$
Actually, the quantization
characteristic surface is the very same for all the field variables, at
variance with Eq.s\ (\ref{7a}), (\ref{7b}) and (\ref{7c}).
It is convenient to introduce
the longitudinal (unphysical) components
of the gauge potential
$
\Gamma_\mu = -\left(\partial_\mu\varphi +
n^*_\mu\lambda\right),
$
\begin{eqnarray}
\Gamma_\mu (x) & =&\int {d^2 k_\perp dk_-\over (2\pi)^{3/2}}
{\theta (k_-)\over \sqrt{|k_\perp|}}
\left\{\left[
-{k_\mu\over |k_\perp|}f(k_\perp ,k_-)\right.\right.\\
& &+\left.\left.n^*_\mu g(k_\perp ,k_-)\right]
e^{-ikx} +\ {\rm h.\ c.}\right\}_{k_+=0},
\label{14}
\end{eqnarray}
whilst
the transversal (physical) components become
$
T_\mu (x)\equiv A_\mu (x) - \Gamma_\mu (x),
$
\begin{eqnarray}
T_\mu (x) & =&
\int {d^2 k_\perp dk_-\over (2\pi)^{3/2}}{\theta (k_-)\over \sqrt{2k_-}}
\varepsilon_\mu^{(\alpha)} (k_\perp ,k_-)\\
& &\times\left\{a_\alpha (k_\perp ,k_-) e^{-ikx} +
a^\dagger_\alpha (k_\perp ,k_-)e^{ikx}\right\}_{k_+=k_\perp^2/2k_-},
\label{15}
\end{eqnarray}
the real polarization vectors being given, e.g., in \cite{6.}.

It is very easy to verify that the canonical light front form algebra
entails
$$
\left[a_\alpha (k_\perp ,k_-),a_\beta^\dagger (p_\perp ,p_-)\right]  =
\delta_{\alpha\beta}\delta^{(2)}(k_\perp -p_\perp)\delta (k_- -p_-),
\label{16a}
$$
\begin{equation}
\left[f(k_\perp ,k_-),g^\dagger (p_\perp ,p_-)\right] =
\delta^{(2)}(k_\perp -p_\perp)\delta (k_- -p_-),
\label{16b}
\end{equation}
all the other commutators vanishing.
Owing to the {\bf LFF} canonical commutation relations (\ref{16b}), it is clear
that
the theory involves
an indefinite metric space of states \cite{4.},\cite{9.} and
the physical Hilbert's subspace
${\cal V}_{{\rm phys}}$
is defined through
$
g(k_\perp ,k_-)\left|{\bf v}\right> = 0,\
\forall\left|{\bf v}\right>\in {\cal V}_{{\rm phys}}.
$
Now,
taking Eq.s\ (\ref{14}-\ref{16b}) into account, Eq.\ (\ref{8}) precisely yields \cite{6.}
the standard Mandelstam-Leibbrandt form
of the
light front form propagator in the anti light cone gauge
$$
\tilde D^+_{\mu\nu} (k)={i\over k^2+i\epsilon}\left\{
-g_{\mu\nu}+{n^*_\mu k_\nu + n^*_\nu k_\mu\over [n^*k]_{\rm ML}}\right\}.
\label{17}
$$
\bigskip
\centerline{{\bf LIGHT-FRONT PURE YANG-MILLS THEORY}}
\medskip
Let us start now from the $SU(N)$-YM Lagrange density
$$
{\cal L}_{{\rm YM}}=-{1\over
4}\left<F_{\mu\nu},F^{\mu\nu}\right>-n^{*\mu}\left<\Lambda, A_\mu\right>,
\label{18}
$$
in which we understand gauge potentials as well as non-Abelian field
strengths
to be $su(N)$-Lie algebra valued fields, $\left<\ ,\ \right>$ being the
inner
product.
In order to quantize the system we shall follow, as it is somewhat customary
in the non-Abelian case, the Hamiltonian path-integral quantization \cite{10.}.
The free (unconstrained) canonical momentum is
$
\pi^- = F_{+-},
$
and we have primary second class constraints
$
\phi^\alpha\equiv\pi^\alpha - F_{-\alpha}=0
$
and primary first class constraints
$
\pi^+ = \pi^\Lambda = 0.
$

The canonical Hamilton density is
\begin{eqnarray}
{\cal H}_{\rm YM}&&\equiv
\left<\pi^-,\partial_+A_-\right>+\left<\pi^\alpha,\partial_+A_\alpha\right>
-{\cal L}_{\rm YM} \\ &&={1\over 2}\left<\pi^-,\pi^-\right>+{1\over
4}\left<F_{\alpha\beta},F^{\alpha\beta}\right>
-\left<A_+,D_\alpha\pi^\alpha+D_-\pi^--\Lambda\right>,
\label{19}
\end{eqnarray}
with
$
D_\mu\equiv {\bf 1}\partial_\mu +ig[A_\mu,\ ].
$
Consequently, we derive the secondary constraints
$
A_+=0,
$
$
D_\alpha\pi^\alpha +D_-\pi^- = \Lambda,$
and
since we have primary second class constraints
$
\phi^\alpha \equiv\pi^\alpha -D_- A_\alpha +D_\alpha A_- =0,
$
satisfying
$$
\left.\left\{\phi^\alpha (x),\phi^\beta(y)\right\}\right|_{x^+=y^+}
=2\delta^{\alpha\beta}D_-(x-y),
\label{20}
$$
with
$$
D_-(x-y)\equiv \left\{{\bf 1}{\partial\over
\partial x^-}+ig[A_-(x),\ ]\right\}\delta^{(3)}({\bf x} -{\bf y}),
\label{21}
$$
then
the Hamiltonian generating functional takes the form
\begin{eqnarray}
&&{\cal Z}[J^\mu]={\cal N}^{-1}\int{\cal D}A_\mu{\cal D}\Lambda{\cal
D}\pi^-{\cal D}\pi^\perp\ \delta\left(\pi^\perp - F_{-\perp}\right)
{\tt det}\Vert D_-\Vert \\
&&\exp i\int d^4 x\left\{\left<\pi^-,\partial_+ A_-\right>
+\left<\pi^\alpha,\partial_+ A_\alpha\right>
-{\cal H}_{\rm YM}+\left<A_\mu, J^\mu\right>\right\}.
\label{24}
\end{eqnarray}
Now the key point: it is well known \cite{11.} that within dimensional
regularization
it turns out that
$$
\left.{\tt det}\Vert D_-\Vert\right|_{\rm dim\ reg}
 = {\tt det}\Vert\partial_-\Vert ,
\label{25}
$$
and
after integration over
$A_+,\pi^\alpha,\pi^-$ and $\Lambda$ one gets
$$
{\cal Z}[J^\perp,J^-]={\cal N^\prime}^{-1}{\tt det}\Vert\partial_-\Vert
\int{\cal D}A_-{\cal
D}A_\perp\
\exp\ i\int d^4 x
$$
$$
\left\{{1\over 2}\left<\partial_+ A_-,\partial_+ A_-\right>
-{1\over 4}
\left<f_{\alpha\beta},f^{\alpha\beta}\right>+\left<f_{-\alpha},\partial_+
A_\alpha\right> \right\}
$$
$$
\times\exp\ i\int d^4 x
\left\{{\cal L}_{\rm Int}
+\left<A_\alpha,J^\alpha\right>+\left<A_- ,J^-\right>\right\},
\label{26}
$$
where it is convenient to separate the Abelian part of the gauge field
strengths $
f_{\mu\nu}\equiv \partial_\mu A_\nu -\partial_\nu A_\mu$, whereas the
interaction Lagrange density
$$
{\cal L}_{\rm Int}(A_-,A_\perp)=
{i\over 2}g\left<[A_\alpha,A_\beta],F_{\alpha\beta}\right>
-ig\left<[A_-,A_\alpha],\partial_+ A_\alpha\right>,
\label{27}
$$
leads to the conventional Feynman's rules for the non-Abelian three-
and four-gluon vertices.
It follows therefrom that the
perturbation theory generating functional
takes the form
$$
{\cal Z}=\exp\left\{ i\int d^4 y\ {\cal L}_{\rm Int}\left<
{\delta\over i\delta J^-(y)},{\delta\over i\delta J^\perp (y)}\right>
\right\}
{\cal Z}_0[J^-,J^\perp],
\label{28}
$$
in which
the free gaussian Abelian generating functional
reads
$$
{\cal Z}_0[J^\perp,J^-]={\cal N}_0^{-1}
\int{\cal D}A_-{\cal
D}A_\perp{\cal D}\pi^-\
\exp\ i\int d^4 x
$$
$$
\left\{-{1\over 2}\left<\pi^-,\pi^-\right>
-{1\over 4}
\left<f_{\alpha\beta},f^{\alpha\beta}\right>+\left<f_{-\alpha},\partial_+
A_\alpha\right> \right\}
$$
\begin{equation}
\times\exp\ i\int d^4 x
\left\{\left<\pi^-,\partial_+A_-\right>
+\left<A_\alpha,J^\alpha)+(A_- ,J^-\right>\right\}.
\label{29}
\end{equation}

We have now to show that the above expression (\ref{29}) for the free generating
functional actually gives rise to the {\bf ML} form of the free gluon
propagator. To this aim, let us perform the change of variables of Eq.\ (\ref{11})
in
the functional integral,
the corresponding Jacobian being
$
{\tt J}=
\Vert\partial_\perp^2\Vert,
$
together with the
sources redefinition
$$
j^\alpha =J^\alpha +2\partial_\alpha\partial_-
\partial_\perp^{-2}* J^-, \quad
\eta =\partial_\alpha J^\alpha+\partial_- J^-.
\label{30}
$$
Then the  free generating functional exactly becomes
$$
{\cal Z}_0 [\eta,j^\perp,J^-]={\cal N}_0^{-1}{\tt
det}\Vert\partial_\perp^2\Vert \int{\cal D}T_\perp{\cal D}\varphi{\cal
D}\lambda\exp\ i\int d^4x $$
$$
\left\{\left<\partial_+T_\alpha,\partial_-T_\alpha\right>-{1\over
2}\left<\partial_\alpha T_\beta,\partial_\alpha T_\beta\right>
+\left<\varphi,\partial_+\partial_\perp^2\lambda\right>\right\}
$$
$$
\times\exp\ i\int d^4x
\left\{\left<T_\alpha, j^\alpha\right> +\left<\varphi,\eta\right>
-\left<\lambda, J^-\right>\right\},
\label{31}
$$
and after choosing standard causal asymptotic conditions
at $x^+\to\pm\infty$ for all the integration field variables \cite{12.},
it is not difficult to prove that
$$
{\cal Z}_0[J^\mu]=\exp\ {1\over 2}\int d^4 x\int d^4 y\ \left<J^\mu(x),
D^+_{\mu\nu}(x-y)J^\nu(y)\right>,
\label{32}
$$
where
$$
D^+_{\mu\nu}(x)=i\int{d^4 k\over (2\pi)^4}
{e^{ikx}\over k^2+i\epsilon}\left\{
-g_{\mu\nu}+{n^*_\mu k_\nu + n^*_\nu k_\mu\over [n^*k]_{\rm ML}}\right\},
\label{33}
$$

In conclusion, we can say that the light front form
perturbative approach in the light cone gauge $A_-=0$ is
still unclear, whereas the
{\bf LFF} perturbative approach in the anti light cone gauge $A_+=0$
is fully consistent.
In particular, since we have seen that the light front form and the instant
form of the generating functional actually coincide - they are nothing but
the
same formal expression, although written using different coordinates systems
and after a suitable rearrangement of the external sources - it immediately
follows that the structure of the counterterms is the very same \cite{4.}. It has
been quite recently noticed
that quark field can also be included into the same approach \cite{6.},\cite{13.}.
Therefore, after fifty years since the original Dirac's
attempt \cite{1.}, we have at our disposal a light front form perturbative
approach
for gauge theories on equal footing as the standard covariant one, what is a
highly non-trivial achievement.
Once a completely consistent light front form formulation has been reached,
we can turn now to attack
non-perturbative {\bf LFF} open issues
such as, e.g., the light front vacuum structure of gauge theories and the
discretized light front quantization.

\end{document}